\documentclass[aps,superscriptaddress, twocolumn,pre,nofootinbib]{revtex4-1}
\usepackage{amsmath,amssymb,amsfonts,amsthm}
\usepackage{tabularx}
\usepackage{graphicx}
\usepackage{array}
\usepackage{comment}
\usepackage{float}
\usepackage{tikz}
\usetikzlibrary{shapes,arrows}
\newcommand{\todo}[1]{}

\begin{document}

\title{Force-Field Functor Theory: \\ Classical Force-Fields which Reproduce Equilibrium Quantum Distributions}

\author{Ryan Babbush}
\affiliation{Department of Chemistry and Chemical Biology, Harvard University, Cambridge MA 02138 USA}

\author{John Parkhill}
\affiliation{Department of Chemistry, The University of Notre Dame, South Bend IN 46556 USA}

\author{Al\'{a}n Aspuru-Guzik}
\affiliation{Department of Chemistry and Chemical Biology, Harvard University, Cambridge MA 02138 USA}

\date{\today }

\begin{abstract}
Feynman and Hibbs were the first to variationally determine an effective potential whose associated classical canonical ensemble approximates the exact quantum partition function. We examine the existence of a map between the local potential and an effective classical potential which matches the \emph{exact} quantum equilibrium density and partition function. The usefulness of such a mapping rests in its ability to readily improve Born-Oppenheimer potentials for use with classical sampling. We show that such a map is unique and must exist. To explore the feasibility of using this result to improve classical molecular mechanics, we numerically produce a map from a library of randomly generated one-dimensional potential/effective potential pairs then evaluate its performance on independent test problems. We also apply the map to simulate liquid \emph{para}-hydrogen, finding that the resulting radial pair distribution functions agree well with path integral Monte Carlo simulations. The surprising accessibility and transferability of the technique suggest a quantitative route to adapting Born-Oppenheimer potentials, with a motivation similar in spirit to the powerful ideas and approximations of density functional theory.
\end{abstract}
\maketitle

The energy and mass scales of chemical motion lie in a regime between quantum and classical mechanics but for reasons of computational complexity, molecular modeling (MM) is largely performed according to Newton's laws. When classical Hamiltonians are chosen to reproduce properties of real material, classical MM is an efficient compromise. An increasing amount of MM uses highly accurate Born-Oppenheimer (BO) potential energy surfaces, which allow one to study complex bond rearrangements where experiment cannot motivate a potential \cite{Wang:2013ul,Car:1985qf}. The BO surface is incompatible with classical statistical mechanics in the sense that we would not expect a classical simulation on the BO surface to reproduce properties of the real material, except in the limit of infinite temperature.

Many approaches already exist to bridge this gap and study quantum equilibrium properties and dynamics: path integral Monte Carlo (PIMC), ring polymer molecular dynamics (RPMD), centroid molecular dynamics (CMD), variational path-integral approximations, discretized path-integral approximations, semi-classical approximations, and colored-noise thermostats \cite{Ceriotti:2011ys,Paesani:2006fr,Nakayama:2003mz,Poulsen:2003ly,Whitlock:1979gf,Liu:2009ve,Jang2001,Fanourgakis2009,Perez2011,Chandler1981}. Most of the these methods involve computational overhead significantly beyond classical mechanics and as they approach exactness their cost rapidly increases.

An alternative philosophy is suggested by density functional theory (DFT) \cite{Hohenberg1964,Mermin1965,Sham1966,Runge1984,Yuen-Zhou2010,Tempel2011}. Following this line of reasoning, three questions arise. Can an equilibrium quantum density be obtained from purely classical mechanics and an effective Hamiltonian? Is the effective Hamiltonian uniquely determined by the physical potential? Can the particle density and free energy be given by such a fictitious system? To all these questions, the answer ``yes'' is implied by the usual recipe for classical force fields that fit experimental data.  This paper examines the uniqueness and existence of a map yielding a classical effective potential given the physical potential.

The bargain of our proposed effective classical potential is similar to that posed by DFT. One sacrifices access to rigorous momentum based-observables and abandons the route to systematic improvement. In exchange, the two properties which are physically guaranteed, the equilibrium particle density and the partition function, are obtained at a cost equivalent to classical sampling but with improved accuracy. As a practical tool, the map is an easy way to transform BO-based force fields into a form which is well-suited for classical sampling. Perhaps the most promising aspect of this mapping would be its scalability which could potentially extend the ability to treat quantum propagation effects to all systems that can be sampled classically. It is even possible that the fictitious trajectories of particles moving on such a potential would, like Kohn-Sham orbitals, have somewhat improved physicality over their classical counterparts, although we will not examine that possibility here.

First, we show the uniqueness of an equilibrium effective potential that gives the exact equilibrium quantum density via classical sampling. Next, we demonstrate that the equilibrium effective potential may be approximated by a linear operator acting on the true potential. Finally, we numerically approximate the map in a rudimentary way, and obtain surprisingly good results and transferability for both one dimensional potentials and a model of liquid \emph{para}-hydrogen. 

\section{Equilibrium effective potential}
In their seminal work on path integral quantum mechanics, Feynman and Hibbs introduced the concept of an effective classical potential that allows for the calculation of quantum partition functions in a seemingly classical fashion \cite{Feynman1965}. In Appendix A, we discuss a connection with the large and fruitful body of research that focuses on the centroid effective potential and density which should not be confused with the equilibrium effective potential that we now define \cite{Feynman1986,Voth1991,Cao1993,Cao1994b,Cao1994a,Cao1994,Martyna1996,Cao1996,Krajewski2001}. We start by considering the equilibrium density matrix,
\begin{equation}
\label{density}
\rho_0 \!\left(q_a, q_b\right) \equiv  \frac{1}{Z}\left\langle q_a \left | e^{-\beta {\cal H}} \right | q_b \right\rangle.
\end{equation}
where ${\cal H}$ is the system Hamiltonian, $\beta$ is the inverse temperature, and $Z$ is the partition function. Feynman showed us that we could connect this expression to the path integral representation of the quantum propagator\footnote{Throughout this paper the variable ``$q$'' refers to a position in the full coordinate space of the system ($q \in \Re^{3 N}$ where $N$ is the number of particles). To distinguish a position variable from a path variable we will use $r\!\left(t\right)$ to represent a particular trajectory.},
\begin{equation}
  \rho_0 \!\left(q_a, q_b\right) = \frac{1}{Z} \int_{r\left(0\right) = q_a}^{r\left(\beta\hbar \right) = q_b}  {\cal D} r\!\left(\tau\right) e^{-{\cal A}\left[r\left(\tau\right)\right]}.
\end{equation}
where the Wick-rotated $(t \rightarrow -i \tau)$ action functional is,
\begin{equation}
{\cal A}\left[r\!\left(\tau\right)\right] = \frac{1}{\hbar}\int_{0}^{\beta \hbar} \textrm{d} \tau \left[\sum_{i=1}^{N}\frac{m_i}{2} \dot{r}_i\!\left(\tau\right)^2+V\!\left(r\!\left(\tau\right)\right)\right].
\label{action}
\end{equation}
By integrating over only closed paths at each coordinate we obtain the scalar equilibrium density,
\begin{equation}
\label{neq}
\eta_0\!\left(q\right) \equiv \frac{1}{Z} \langle q | \rho_0 | q\rangle =  \frac{1}{Z}\oint_{r\left(0\right) = q}^{r\left(\beta\hbar \right) = q}  {\cal D} r\!\left(\tau\right) e^{-{\cal A}\left[r\left(\tau\right)\right]}.
\end{equation}
Finally, we define the partition function as a normalization factor which is obtained by integrating over $q$,
\begin{equation}
\label{Z}
Z  \equiv \textrm{Tr}\left[e^{-\beta {\cal H}}\right] =  \int_{-\infty}^{\infty} \!\!\textrm{d} q \oint_{r\left(0\right) = q}^{r\left(\beta\hbar \right) = q}  {\cal D} r\!\left(\tau\right) e^{-{\cal A}\left[r\left(\tau\right)\right]}.
\end{equation}

We are now in a position to define an equilibrium effective potential, which encapsulates knowledge of the physical quantum density into a form amenable to classical sampling. We choose the equilibrium effective potential, $W\!\left(q\right)$ such that,
\begin{eqnarray}
\label{nclass}
\eta_0\!\left(q\right) & \equiv & \frac{1}{Z} e^{-\beta W\left(q\right)} \\
\label{W}
W\!\left(q\right) & \equiv & -\frac{1}{\beta}\log\left[\oint_{r\left(0\right) = q}^{r\left(\beta\hbar \right) = q}  {\cal D} r\!\left(\tau\right) e^{-{\cal A}\left[r\left(\tau\right)\right]}\right].
\end{eqnarray}
Note that this definition associates the Boltzmann factor, $e^{-\beta W\left(q\right)}$, with the \emph{unnormalized} density. Because $\eta_0\!\left(q\right)$ must integrate to unity, this allows us to easily recover the partition function and corresponding quantum Helmholtz free energy, $A$, with the classical integral,
\begin{equation}
\label{Zclass}
\int_{-\infty}^{\infty} \!\textrm{d}q \, e^{-\beta W\left(q\right)} = Z \int_{-\infty}^{\infty} \!\textrm{d}q\, \eta_0\!\left(q\right) = Z \equiv e^{-\beta A}.
\end{equation}

Using Eq.~\ref{W}, one can exactly calculate the equilibrium effective potential whenever one can evaluate the path integral. Unfortunately that is usually numerically intractable. Thus, it is useful to wonder if a $\emph{unique}$ map exists between any potential $V\!\left(q\right)$ and $W\!\left(q\right)$ under the conditions of a fixed ensemble. If one could easily evaluate the map one could transferably adapt BO potentials to give physical results in classical simulations. Since this mapping is a functor\footnote{A functor differs from a functional in that a functor maps one vector space to another whereas a functional maps a vector space to a scalar. In this context, ``operator'' is a more common term than ``functor'' but we prefer to call this ``force-field functor theory'' to evoke the connection with DFT.} which gives an effective force-field we refer to the map as the ``force-field functor'' and denote it with the symbol ${\cal F}$.

\section{Uniqueness and Existence}
Our first step towards developing a theory of force-field functors is to show that the proposed mapping,  ${\cal F} \!\left[V\!\left(q\right)\right] \rightarrow W\!\left(q\right)$, exists and is unique. This proof begins in Part A of the current section in which we argue that no two $V\!\left(q\right)$ lead to the same quantum equilibrium density $\eta_0\!\left(q\right)$, which exists by Eqs.~\ref{action} and~\ref{neq}. To show this we take inspiration from Mermin's extension of the Hohenberg-Kohn theorem for finite temperatures and use the quantum Bogoliubov inequality to construct a proof by contradiction \cite{Mermin1965}. For any potential without hard-shell interactions, the density is always given by a Boltzmann factor of the potential as in Eq.~\ref{nclass}; thus, the equilibrium effective potential exists for any physically-relevant quantum potential. In Part B of the current section, we make a similar argument to prove that there is a one-to-one map between classical equilibrium density and classical potential \cite{Chayes:1984dq}. Since the effective potential is chosen to be the classical potential associated with the quantum density, these results directly imply that the map between physical potential and effective potential must be unique.
\begin{figure}[h]
\label{diagram}
\centering
\begin{tikzpicture}[>=latex]
    \node[draw, circle] at (0,0) (V) {$V\!\left(q\right)$};
    \node[draw,ellipse] at (3.5,0) (n) {$\eta_Q\!\left(q\right) \equiv \eta_0\!\left(q\right)$};
    \node[draw,circle] at (7,0) (W) {$W\!\left(q\right)$};
    \draw[->, bend left] (V) to node [above] {Exists (Eq.~\ref{neq})} (n);
    \draw[->, bend right] (V) to node [below] {Unique (Section IIA)} (n);
    \draw[->, bend left] (n) to node [above] {Exists (Eq.~\ref{W})} (W);
    \draw[->, bend right] (n) to node [below] {Unique (Section IIB)} (W);
\end{tikzpicture}
\caption{Morphism depicting uniqueness and existence of mappings between the physical potential, $V\!\left(q\right)$, the equilibrium effective potential, $W\!\left(q\right)$, and the associated quantum and classical equilibrium densities, $\eta_Q\!\left(q\right)$ and $\eta_{0}\!\left(q\right)$, respectively. This establishes the existence of a mapping, ${\cal F}$, which uniquely determines the equilibrium effective potential.}
\end{figure}
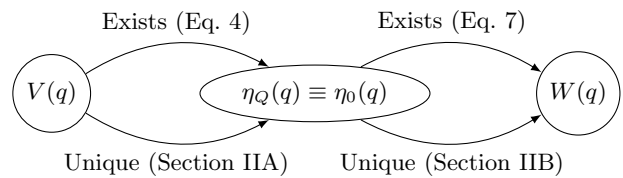

\subsection{Uniqueness of quantum density}

Both steps in this proof take the form of \emph{reductio ad absurdum} arguments based on the uniqueness of an ensemble which minimizes the free energy of a canonical system. In the Appendix B we show that,
\begin{equation}
A\left[\rho\right] > A\left[\rho_0\right], \, \, \, \rho \neq \rho_0
\end{equation}
where $A$ is the quantum Helmholtz free energy,
\begin{equation}
A\left[\rho\right] = \textrm{Tr}\left[\rho\left({\cal H} + \frac{1}{\beta}\log\left[\rho\right]\right)\right],
\end{equation}
which is minimum when $\rho$ is equal to the quantum equilibrium density matrix $\rho_0$ associated with the Hamiltonian, ${\cal H} = T + V\!\left(q\right)$. With this in mind, suppose that there were another potential $\widetilde{V}\!\left(q\right)$ that led to the same density $\eta_0\!\left(q\right)$. Denote the Hamiltonian, canonical density matrix and Helmholtz free energy associated with $\widetilde{V}\!\left(q\right)$ by $\widetilde{{\cal H}}$, $\widetilde{\rho}_0$, and $\widetilde{A}$. Since $\widetilde{V}\!\left(q\right) \neq V\!\left(q\right)$ and $\widetilde{\rho_0} \neq \rho_0$\footnote{That the corresponding equilibrium density matrices are not equal is obvious in Eq.~\ref{density}.} we can write
\begin{eqnarray}
\widetilde{A} & = & \textrm{Tr}\left[\widetilde{\rho_0}\left(\widetilde{{\cal H}}+\frac{1}{\beta}\log\left[\widetilde{\rho_0}\right]\right)\right]\\
& < & \textrm{Tr}\left[\rho_0\left(\widetilde{{\cal H}} + \frac{1}{\beta}\log\left[\rho_0\right]\right)\right]\nonumber\\
& = & A + \textrm{Tr}\left[\rho_0 \widetilde{V}\!\left(q\right) - \rho_0 V\!\left(q\right)\right]\nonumber .
\end{eqnarray}
Using the definition of the quantum equilibrium particle density,\footnote{Recall that $q \in \Re^{3 N}$ so, $|q\rangle = \prod_{i=1}^{3N} |q_i\rangle$.}
\begin{equation}
\eta_0\!\left(q\right) \equiv \textrm{Tr}\left[\rho_0 |q\rangle \langle q|\right],
\end{equation}
we see that,
\begin{equation}
\widetilde{A} < A + \int\textrm{d}q\left[\widetilde{V}\!\left(q\right) - V\!\left(q\right)\right]\eta_0\!\left(q\right).
\end{equation}
But we see that this relation is still true if we interchange over-scored variables,
\begin{equation}
A < \widetilde{A} + \int\textrm{d}q\left[V\!\left(q\right) - \widetilde{V}\!\left(q\right)\right]\eta_0\!\left(q\right).
\end{equation}
This leads to the contradiction,
\begin{equation}
A + \widetilde{A} < \widetilde{A} + A.
\end{equation}
and therefore only one $V\!\left(q\right)$ can result in a given $\eta_0\!\left(q\right)$. This proves that $V\!\left(q\right)$ uniquely determines $\eta_0\!\left(q\right)$. Next, we show that the only potential which can reproduce the quantum density with classical sampling is the equilibrium effective potential.

\subsection{Uniqueness of the effective potential}

Eq.~\ref{W} shows the existence the equilibrium effective potential, $W(q)$. It remains to be shown that $W(q)$ is the only such potential which will reproduce the quantum density, which is to say that ${\cal F}$ is completely unique. The classical Bogoliubov inequality states that,
\begin{equation}
A\left[\tilde{\eta_0}\!\left(q\right)\right] > A\left[\eta_0\!\left(q\right)\right], \, \, \, \tilde{\eta_0}\!\left(q\right) \neq \eta_0\!\left(q\right)
\end{equation}
where $A$ is the classical Helmholtz free energy,
\begin{eqnarray}
A\left[\eta_0\!\left(q\right)\right] & = & E\left[\eta_0\!\left(q\right)\right]-\frac{1}{\beta}S\left[\eta_0\!\left(q\right)\right]\\
& = & \int \textrm{d}q \,\eta_0\!\left(q\right) W\!\left(q\right) + \frac{1}{\beta}\int \textrm{d}q \,\eta_0\!\left(q\right) \log\left[\eta_0\!\left(q\right)\right]\nonumber
\end{eqnarray}
which is minimum when $\eta_0\!\left(r\right)$ is equal to the classical equilibrium density in the presence of $W\!\left(q\right)$. For completeness, this result is also proved in Appendix C. With this in mind, suppose that there were two effective potentials, $\widetilde{W}\!\left(q\right)$ and $W\!\left(q\right)$ that led to the same density. Then,
\begin{eqnarray}
\widetilde{A} & = & \int \textrm{d}q \,\eta_0\!\left(q\right) \widetilde{W}\!\left(q\right) + \frac{1}{\beta}\int \textrm{d}q \,\eta_0\!\left(q\right) \log\left[\eta_0\!\left(q\right)\right]\\
& < & \int \textrm{d}q \,\eta_0\!\left(q\right) W\!\left(q\right) + \frac{1}{\beta}\int \textrm{d}q \,\eta_0\!\left(q\right) \log\left[\eta_0\!\left(q\right)\right]\nonumber\\
& = & A + \int \textrm{d}q \,\eta_0\!\left(q\right) \left[\widetilde{W}\!\left(q\right)-W\!\left(q\right)\right].\nonumber
\end{eqnarray}
So we see that,
\begin{equation}
\widetilde{A} < A + \int \textrm{d}q \,\eta_0\!\left(q\right) \left[\widetilde{W}\!\left(q\right)-W\!\left(q\right)\right].
\end{equation}
If we interchanged all over-scored quantities, we would also find the following,
\begin{equation}
A < \widetilde{A} + \int \textrm{d}q \,\eta_0\!\left(q\right) \left[W\!\left(q\right)-\widetilde{W}\!\left(q\right)\right].
\end{equation}
Adding these equations together leads to the result,
\begin{equation}
\widetilde{A} + A < A + \widetilde{A}.
\end{equation}
Thus, we see that no two $W\!\left(q\right)$ lead to the same $\eta_0\!\left(q\right)$.

Because the physical potential $V\!\left(q\right)$ uniquely determines the quantum equilibrium density $\eta_0\!\left(q\right)$, and the quantum equilibrium density uniquely determines the equilibrium effective potential $W\!\left(q\right)$, we see that the map, ${\cal F} \!\left[V\!\left(q\right)\right] \rightarrow W\!\left(q\right)$ must be completely unique.

\section{Approximate Linearity}

The results of the above section establish the possibility of reversing $\cal{F}$ by modeling pairs of $V(q)$ and $W(q)$ generated via the exact path-integral. However the concept of ${\cal F}$ is not useful unless we have good reason to suspect that ${\cal F}$ or a useful approximation to ${\cal F}$ will be easy to obtain and evaluate numerically. In this section, we analyze the approximation of $\cal{F}$ as a linear functor which is straightforwards to construct numerically and because of its separability, applicable to systems of arbitrary dimensionality. 

We begin by rewriting Eq.~\ref{neq} and Eq.~\ref{nclass},
\begin{equation}
e^{-\beta W\left(q\right)} \equiv \oint_{r\left(0\right) = q}^{r\left(\beta\hbar \right) = q}  {\cal D}r\!\left(\tau\right) e^{-{\cal A}\left[r\left(\tau\right)\right]}
\end{equation}
and introduce several definitions which break apart the action term into a kinetic part and a potential part,
\begin{eqnarray}
{\cal U}\!\left[r\!\left(\tau\right)\right] & \equiv &  \frac{1}{\hbar} \oint_{0}^{\beta \hbar} \! \textrm{d} \tau \,\left[V\!\left(q\right) - V\!\left(r\!\left(\tau\right)\right)\right]\\
{\cal T}\!\left[r\!\left(\tau\right)\right] & \equiv & \exp\left[-\frac{1}{2 \hbar} \int_{0}^{\beta \hbar} \! \textrm{d} \tau \, \sum_{i=1}^{N} m_i \, \dot{r}_i\!\left(t\right)^2 \right]\\
\label{simple}
Z_{\cal T} &  \equiv & \int_{-\infty}^{\infty} \!\textrm{d}q\, \oint_{r\left(0\right) = q}^{r\left(\beta\hbar \right) = q} \! {\cal D} r\!\left(\tau\right) {\cal T}\!\left[r\!\left(\tau\right)\right].
\end{eqnarray}
We now employ a notation due to Feynman and Hibbs, for the equilibrium average of a path functional weighted by ${\cal T}$ and normalized by $Z_{\cal T}$, ``$\langle \rangle$'' \cite{Feynman1965}. This allows us to write a concise, exact expression for $W\!\left(q\right)$:
\begin{equation}
e^{-\beta W\left(q\right)} =   Z_{\cal T}\, e^{-\beta V\left(q\right)} \langle e^{{\cal U}\left[r\left(\tau\right)\right]} \rangle .
\end{equation}
Jensen's inequality tells us that that, $\langle e^f \rangle \geq e^{\langle f\rangle}$ with an error on the order of the variance of $f$. This simplifies the path integral and introduces error that is second order at worst in the weighted path functional average,
\begin{equation}
\langle e^{{\cal U}\left[r\left(\tau\right)\right]} \rangle = e^{\langle{\cal U}\left[r\left(\tau\right)\right]\rangle } + {\cal O}\!\left[ \langle {\cal U}\!\left[r\!\left(\tau\right)\right] \rangle^2 - \langle {\cal U}\!\left[r\!\left(\tau\right)\right]^2 \rangle \right]
\end{equation}
\begin{equation}
e^{-\beta W\left(q\right)} \approx Z_{\cal T}\, e^{-\beta V\left(q\right)} e^{\langle {\cal U}\left[r\left(\tau\right)\right] \rangle}.\end{equation}
Because any potential is unique only up to a constant, we can use properties of logarithms to remove $Z_{\cal T}$, since it does not depend on $q$ or $V\!\left(q\right)$, to write
\begin{equation}
\label{approx}
W\!\left(q\right) \approx V\!\left(q\right) - \frac{1}{\beta} \langle {\cal U}\left[r\!\left(\tau\right)\right] \rangle
\end{equation}
with corrections on the order of ${\cal U}^2$. We also see from this that the equilibrium effective potential is a temperature dependent correction to the true potential. ${\cal U}\left[r\!\left(\tau\right)\right] $ is clearly a linear functional of $V\!\left(q\right)$ and $\langle {\cal U}\left[r\!\left(\tau\right)\right] \rangle$ is clearly a linear functor of ${\cal U}\left[r\!\left(\tau\right)\right]$,
\begin{equation}
\label{U}
\langle {\cal U}\!\left[r\!\left(\tau\right)\right] \rangle = \frac{1}{Z_{\cal T}}   \oint_{r\left(0\right) = q}^{r\left(\beta\hbar \right) = q} \! {\cal D}r\!\left(\tau\right) {\cal T}\!\left[r\!\left(\tau\right)\right] \, {\cal U}\!\left[r\!\left(\tau\right)\right].
\end{equation}

In the multi-dimensional case, the path integral couples all $3N$ modes of $q$, making the exact ${\cal F}$ a very complicated object which embeds all-orders of quantum many body effects between these modes. However, our analysis suggests a linear approximation which conserves the locality of the original potential. With this approximation we can separate the integral in Eq.~\ref{U} into each individual interaction order of the potential and see that the path integral does not multiply these terms;  the pairwise interactions remain pairwise, the three-mode interactions are mapped by $\cal{F}$ onto three-mode interactions, etc.

Approximate separability of this mapping is one of the key differences between our method and approaches such as Feynman-Kleinert, which introduces higher ordered many-body terms into the effective potential, or RPMD, which avoids the issue at the cost of introducing ancilla particles. Our $\cal{F}$ can be imagined as a Gaussian smearing of $V(q)$ to first approximation. It is reasonable to suspect that the non-separable many body couplings would be blurred to a high order such that the many-body expansion of the equilibrium effective potential might terminate faster than the many-body expansion of the uncorrected physical potential. This agrees with the empirical observation that tunneling effects stabilize pairwise interactions more than higher-ordered interactions.

\section{Numerical tests}

It is far from obvious that a transferable map between $V\!\left(q\right)$ and $W\!\left(q\right)$ can be practically obtained and usefully accurate. Instead of calling upon the most sophisticated procedures we can implement to solve the problem, we take \emph{the simplest} approach to developing and testing our approximation to ${\cal F}$ so that our results are designed to be a worst-case, upper-bound on the error that leaves room for optimism. Approaches such as machine learning could be employed in future work \cite{Snyder:2012zr}. We approximate ${\cal F}$ as a linear map (a matrix) acting on our potentials vectorized into coefficients of Legendre polynomials. The entries of this matrix are determined by simple least-squares on a randomly generated training set of 1,000 one-dimensional potentials and their corresponding effective potentials chosen by randomly choosing Legendre coefficients with the only constraint being that the classical densities vanish at their boundaries.

Effective potentials were calculated using Eq.~\ref{W} with densities obtained from the efficient real-space discrete variable representation (DVR) of the path integral \cite{Thirumalai1983}. We examine how this ${\cal F}$ performs on instances of other random potentials not included within its training set and then apply it to the Silvera-Goldman pair potential for liquid \emph{para}-hydrogen \cite{Silvera1978,Nakayama2003,Poulsen2004,Hone2004,Miller2005}. Using the resulting effective potential, a classical Monte Carlo simulation was performed to give us radial pair distribution functions in agreement with results from PIMC at a fraction of the computational cost.

\subsection{Obtaining the linear functor}
In order to obtain the simplest possible ${\cal F}$ we model the linear transformation as a matrix. This requires that we treat the physical potential and effective potential as vectors in some basis of real-valued functions. Because force-fields are often chosen for the speed with which they can be evaluated, it seems natural to use a polynomial basis. Legendre polynomials evaluated on a fixed domain of $\left [ -1, 1\right]$ were chosen for their orthogonality and historical usefulness in fitting potentials.

Consider the short time Trotterization of the path-integral, which we use to generate exact quantum densities for our test sets \cite{Thirumalai1983}. The short-time propagator effectively acts as a Gaussian which blurs out the density with a variance that depends exactly on the inverse of the square root of the the mass times the temperature. This factor which determines the ``quantumness'' of the system is proportional to the thermal de Broglie wavelength, $\Lambda = \hbar \sqrt{2\pi \beta / m}$  \cite{Yonetani2003, Georgescu2013}. Because we wish to calculate the deformation of a potential vector evaluated on a fixed domain, the parameter which characterizes our map must depend on the ratio between the thermal de Broglie wavelength and the potential length-scale, $Q = \Lambda / L$ where $L$ is the potential length-scale.

In order to obtain a linear functor capable of transforming a one-dimensional potential at fixed $Q$ into another one-dimensional potential at fixed $Q$ we randomly generated pairs of potential vectors and their corresponding effective potential vectors. These vectors were in a Legendre polynomial basis of order $B$ and the vector elements of the classical potential (i.e. basis coefficients) were drawn from a flat distribution between $-10/\beta$ and $10/\beta$. The corresponding effective potential vectors were calculated by evaluating the classical potential vectors as Legendre polynomials on the fixed domain, passing the scalar potential and $Q$ to the aforementioned DVR routine which yields a scalar quantum density, and finally fitting the negative logarithm of that density divided by $\beta$ to a vector of Legendre polynomials in accordance with Eq.~\ref{W}. Having done this, the goal is to find a matrix ${\cal F} \in B \times B$ such that, ${\cal F} \vec{V} \approx \vec{W}$. We chose to perform a Levenberg-Marquart L2 optimization to determine the elements of this matrix \cite{Levenberg1944}. Our residual was defined as the concatenation of the difference vectors, ${\cal F} \vec{V}_i - \vec{W}_i$ for all $N$ physical potential / effective potential pairs in the randomly generated set. 
\begin{figure}[h]
\centering
\includegraphics[scale=1]{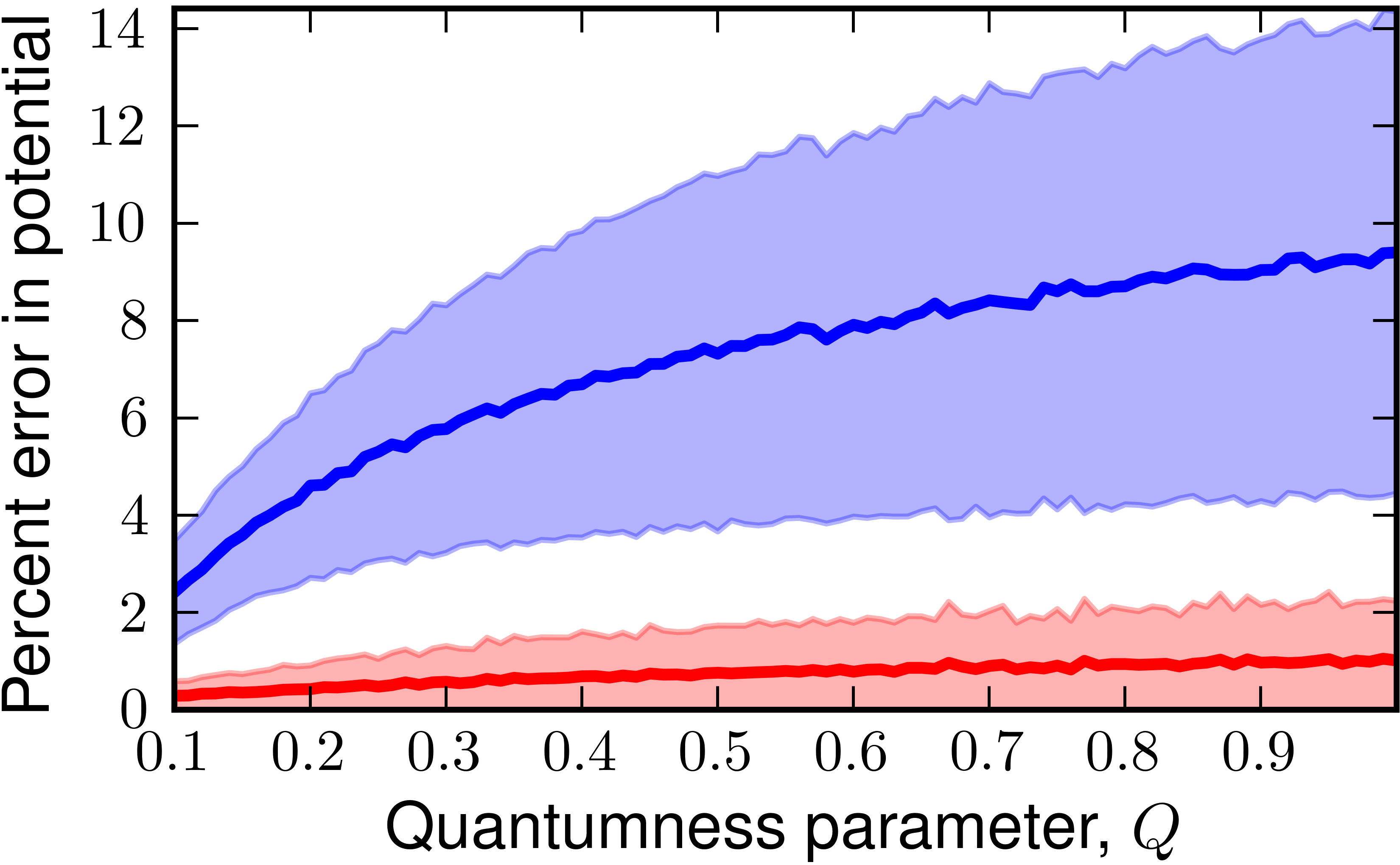}
\includegraphics[scale=1]{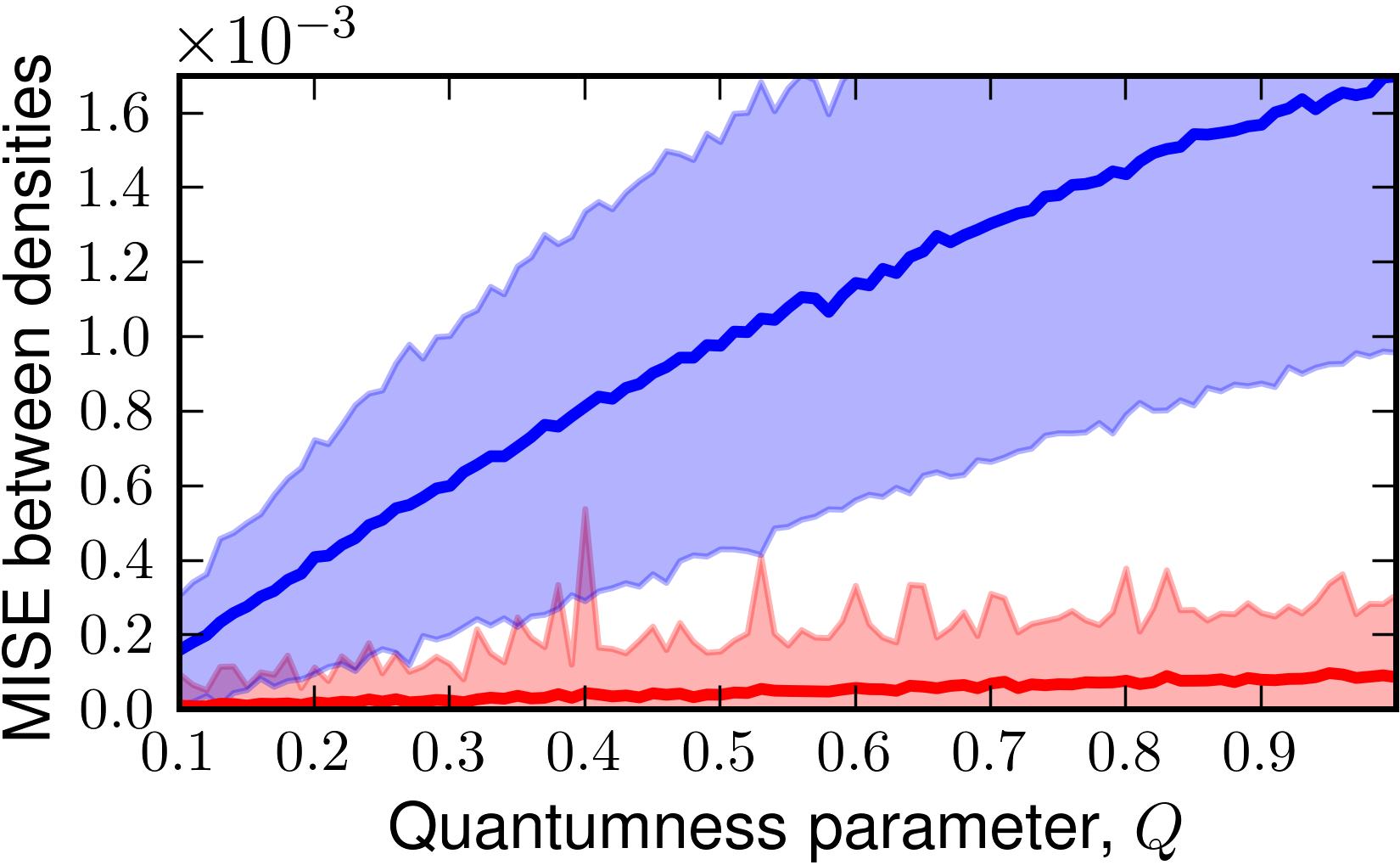}
\caption{Top: plot of the percent error in potential energy of a classical simulation with the classical potential (blue) and ${\cal F}$ generated distribution (red) against $Q$. Bottom: plot of mean integrated squared error (MISE) from the exact quantum density for classical density (blue) and ${\cal F}$ generated density (red) against $Q$. Each point is the mean of these errors on 1,000 random potentials with 50 basis functions.}
\vspace{-10pt}
\label{fig:Perror}
\end{figure}

\subsection{Performance analysis}

\begin{figure*}[htpb]
\centering
\includegraphics[scale=1]{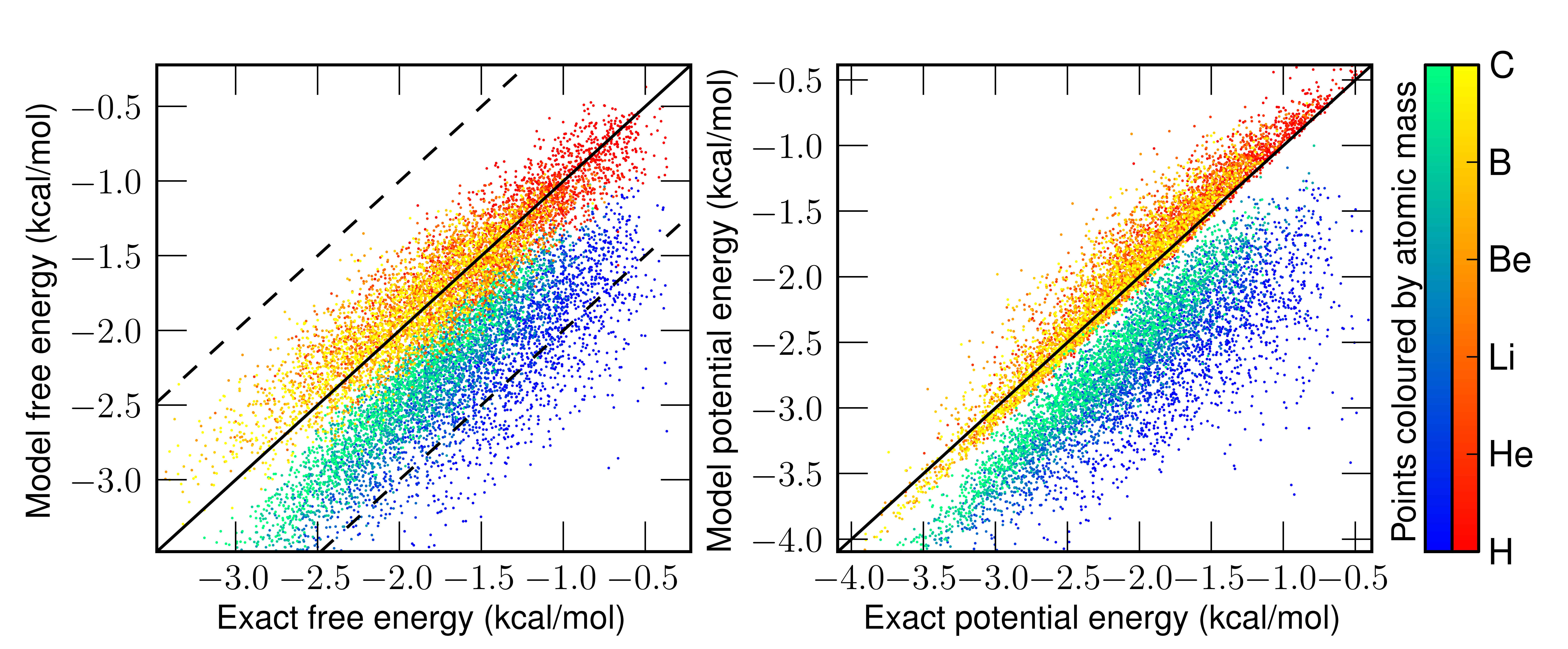}
\caption{Left: correlation of classical (blue-green) and ${\cal F}$ generated (red-yellow) free energy with exact free energy. Dotted lines enclose the chemically accurate region of within one kcal/mol. \emph{In more than 97\% of instances}, our map is more accurate than the classical treatment. Right: correlation of classical and ${\cal F}$ generated potential energy with exact potential energy. Colour brightness indicates the mass used in setting the $Q$ value at $25$K. As mass increases, classical simulations better approximate the energy. Data consists of 1,000 cross-validating potentials at each of the six masses shown on the colourbar.}
\label{fig:dust}
 \vspace{-5pt}
\end{figure*}

The linear approximation to $\cal{F}$ appears to work quite well for even fairly large values of $Q$. As we can see in Figure \ref{fig:Perror}, the errors on an independent test set from the linear ${\cal F}$ generated $W(q)$ are minimal and significantly better than the classical predictions, especially in strongly quantum regimes. Even the deviation from the exact answer is improved relative to simulations which employ the uncorrected physical potential. For both simulations the error goes to zero as $Q$ goes to zero - a consequence of classical correspondence. As one might expect as $Q$ is increased, predictions given by both classical and ${\cal F}$ generated distributions deviate more significantly from the exact answer. In the $W(q)$ simulations these errors are entirely due to the linearity of $\cal{F}$. Another view of the the performance of the linear functor is given in Figure \ref{fig:dust}. When temperature and length are fixed, mass is a reasonable predictor of the performance of both $W(q)$ and $V(q)$ simulations. For low masses, the classical treatment often misses the quantum free energy by as much as a kcal/mol (chemical accuracy). Having characterized the error of assuming linearity we turn to separability.

We apply our linear ${\cal F}$, trained at 14K and 25K with sets of 1,000 potentials, to the Silvera-Goldman potential, which is perhaps the most common potential used to simulate liquid hydrogen with path integral methods \cite{Silvera1978,Nakayama2003,Poulsen2004,Hone2004,Miller2005}. We then performed a classical Monte Carlo simulation on the potential mapped at 25K and the potential mapped at 14K, using 150 molecules in a cubic cell with periodic boundary conditions and one million steps. Cell size was fixed by densities from the literature \cite{Nakayama2003}.
\begin{figure}[H]
\vspace{-5pt}
\centering
\includegraphics[scale=1]{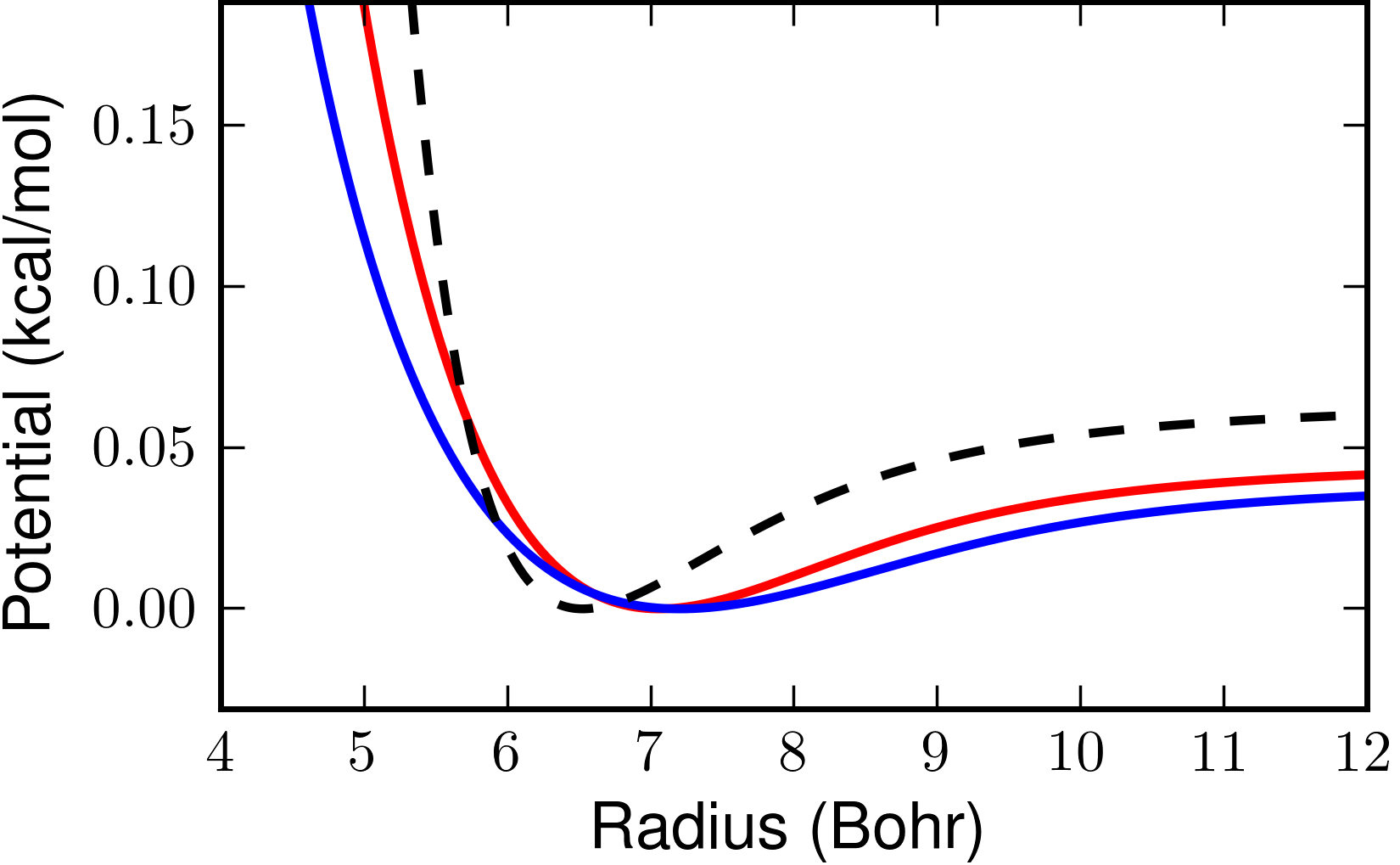}
\caption{The dashed black line above shows the classical Silvera-Goldman potential in the region of interest for our problem. The red line is the effective potential obtained with our linear ${\cal F}$ at 25K and the blue line is at 14K.}
\end{figure}

\begin{figure}[H]
\centering
\vspace{-10pt}
 \hspace{-15pt}
\includegraphics[scale=1]{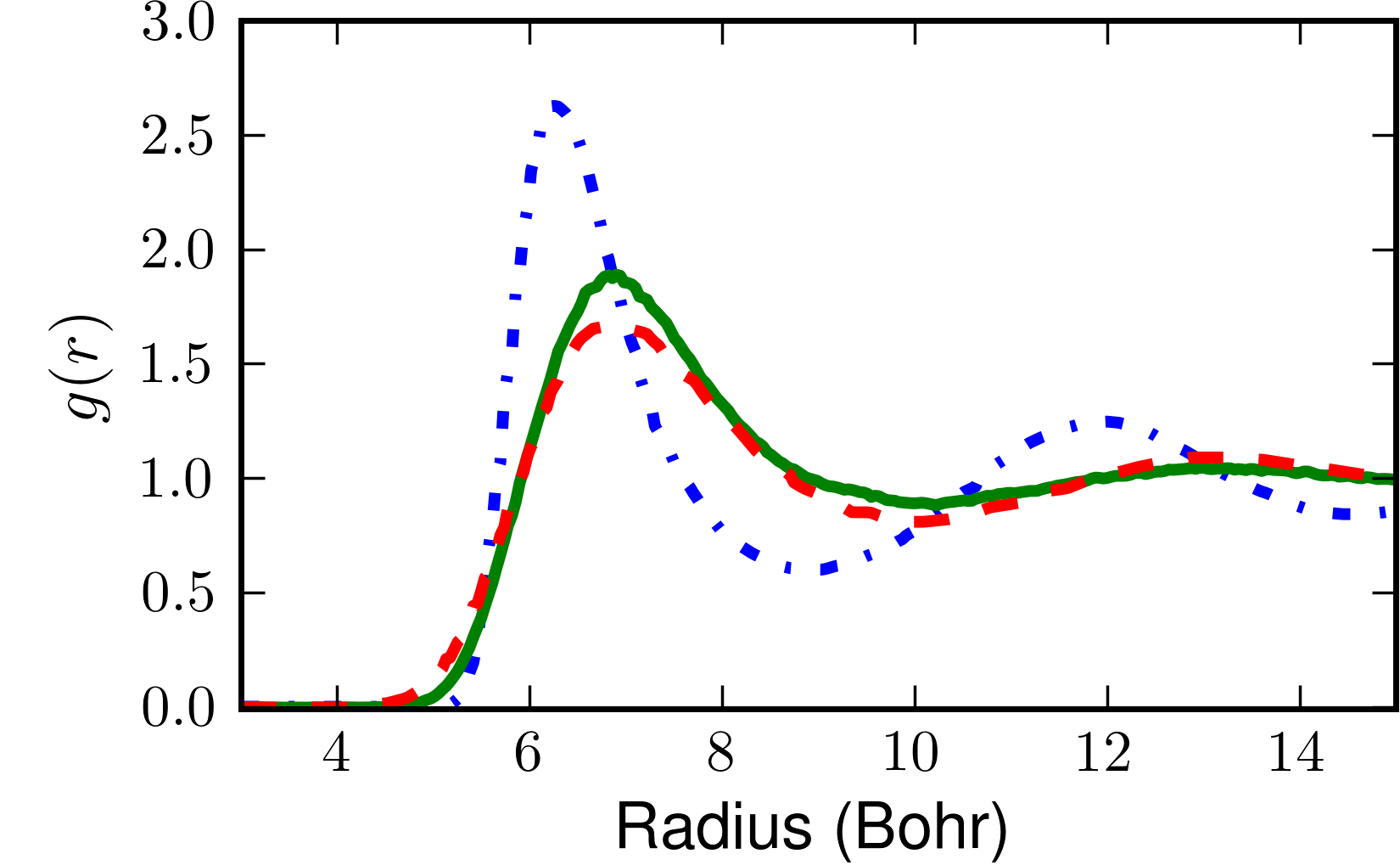}\\
 \hspace{-15pt}
\includegraphics[scale=1]{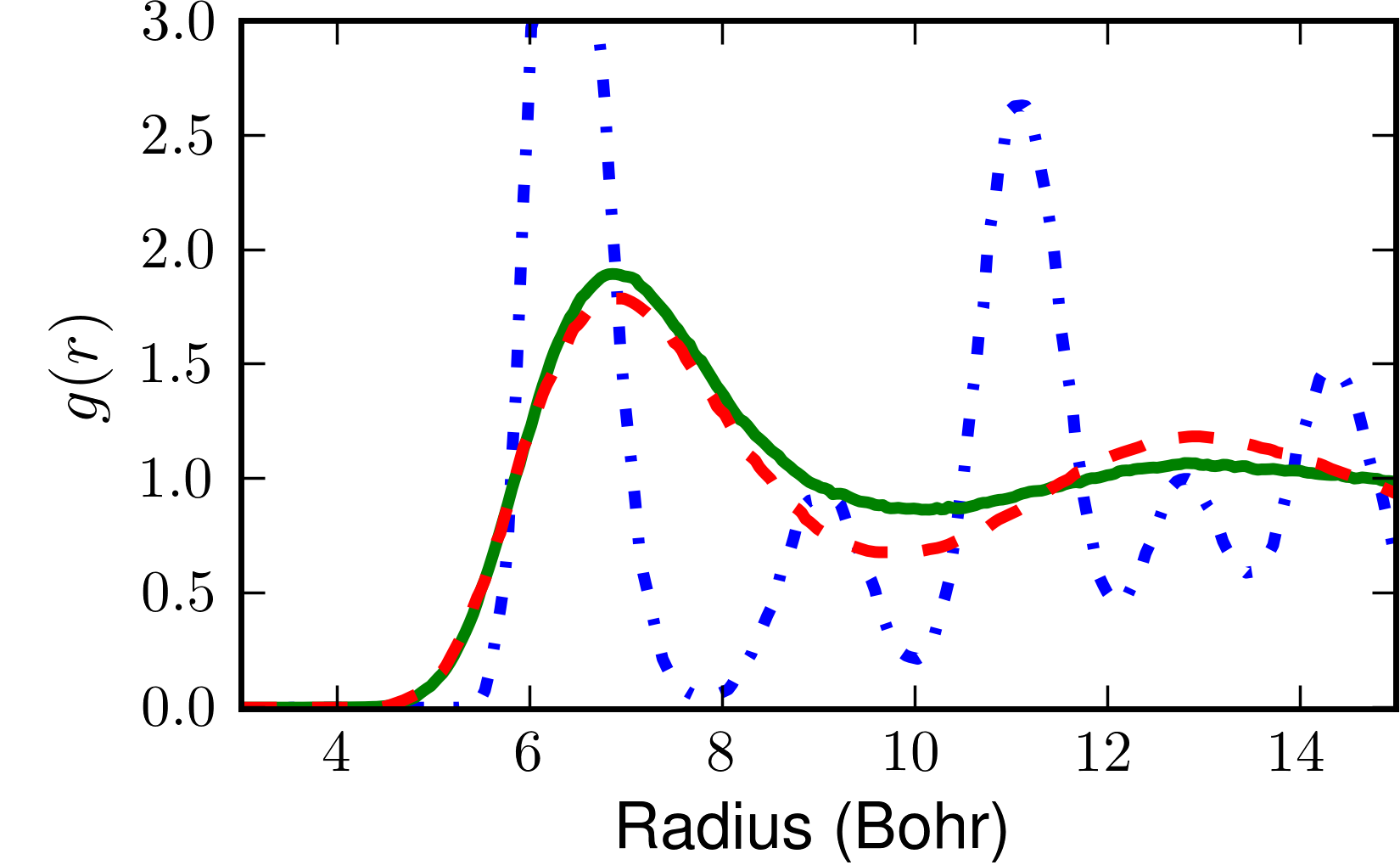}\\
 \vspace{-5pt}
\caption{The top box shows radial pair distribution functions at 25K and the bottom box shows radial pair distribution functions at 14K. The blue (dotted-dashed) curve is for the classical liquid without correction for quantum effects. The green curve (solid) shows the result of classical Monte Carlo sampling on the effective potential obtained with our linear ${\cal F}$. The red curve (dashed) shows PIMC results \cite{Nakayama2003}. Even this simple ${\cal F}$ is a major improvement over the classical potential.}
\label{fig:rdfs}
\end{figure}

The resulting radial distribution functions, $g(r)$ are shown in Figure \ref{fig:rdfs}. The differences between the $W(q)$ generated $g(r)$ and the PIMC results are presumably due to the assumption of separability. Slight over-structuring of $g(r)$ at the first shell results from neglect of the 3-body components of the exact $W(q)$. Remarkably, this over-structuring appears to decrease, with temperature lending credence to the idea that many-body effects in $W(q)$ are largely blurred-out by the smearing which the low orders of $\cal{F}$ perform on the potential.  At both temperatures the errors of these approximations are quite reasonable and although the classical system undergoes a non-physical transition to a solid between 25K and 14K, the model of the present work remains correct.

\section{Conclusion}

We have shown that for each physical potential, there is a unique effective potential which reproduces the quantum density and free energy when sampled with classical statistics. Other properties of a classical simulation of the effective Hamiltonian are not designed to approximate reality by the mapping, but the effective potential may be advantageous to the status quo: classical simulation on a Born-Oppenheimer surface. In this paper we have shown that the implied mapping between the physical and effective potential, ${\cal F}$, can be made concrete to a useful degree of accuracy. A simple linear model for ${\cal F}$ improves on the physical potential systematically over a broad range conditions. Even under the assumption of separability and without any exponential functions in our training set, our model for ${\cal F}$ adequately describes the density of a popular \emph{para}-hydrogen model at exactly the cost of the corresponding classical simulation. Nonlinear models for ${\cal F}$ and expressions which do not assume complete separability are likely to improve on these results and produce even more accurate transferable recipes for digesting Born-Oppenheimer potentials. Ultimately, we hope that force-field functors will provide a scalable methodology for including quantum propagation effects in systems that are intractable for exact methods, such as protein force-fields.

\section{Appendix}
\subsection{Quantum densities from classical sampling}

In practice, path integral expressions are analytically intractable except in a few cases. Feynman proposed to simplify Eq.~\ref{Z} by changing from an integral over all closed paths that start and end at point $q$ to an integral over all closed paths that have an average value equal to the path centroid $\bar{r}$,
\begin{equation}
\bar{r} = \frac{1}{\beta \hbar}\int_0^{\beta\hbar} \!\textrm{d}\tau\, r\!\left(\tau\right).
\end{equation}
So that we only integrate over each closed path once, we must change our expression for the partition function to only calculate paths that match the centroid,
\begin{eqnarray}
\label{Zc}
Z  & = & \int_{-\infty}^{\infty} \!\!\textrm{d} q \oint {\cal D}r\!\left(\tau\right) \delta\left[q - \bar{r}\right] e^{-{\cal A}\left[r\left(\tau\right)\right]}\\
& = & \oint {\cal D}r\!\left(\tau\right) e^{-{\cal A}\left[r\left(\tau\right)\right]}\nonumber .
\end{eqnarray}
While the partition functions given by Eq.~\ref{Z} and Eq.~\ref{Zc} are exactly equal, the two expressions are associated with subtly different scalar density functions. Eq.~\ref{Z} is associated with the true equilibrium density in Eq.~\ref{neq} and Eq.~\ref{Zc} is associated with the path centroid density,
\begin{equation}
\eta_c\!\left(q\right) = \frac{1}{Z}\oint {\cal D}r\!\left(\tau\right) \delta\left[q - \bar{r}\right] e^{-{\cal A}\left[r\left(\tau\right)\right]}.
\end{equation}
The Dirac delta function in this equation enforces the requirement that integrating the Boltzmann factor associated with this density over the path centroid, $\bar{r}$, will result in exactly the path integral expression for the quantum partition function \cite{Jang1999}. The centroid density plays a prominent role in CMD and Feynman-Kleinert methods but does not apply to force-field functor theory.

\subsection{Proof of quantum Bogoliubov inequality}

The quantum Bogoliubov inequality is proved for the grand canonical ensemble in the Appendix of \cite{Mermin1965}. We adapt this proof for the canonical ensemble, in the interest of completeness, to show that for all positive definite $\rho$ with unit trace,
\begin{equation}
A\left[\rho\right] > A\left[\rho_0\right], \, \, \, \rho \neq \rho_0
\end{equation}
if $A$ is the quantum Helmholtz free energy of the canonical ensemble,
\begin{equation}
A\left[\rho\right] = \textrm{Tr}\left[\rho\left({\cal H} + \frac{1}{\beta}\log\left[\rho\right]\right)\right],
\end{equation}
which is minimum only when $\rho$ is equal to the quantum equilibrium density matrix $\rho_0$ associated with the Hamiltonian, ${\cal H} = T + V\!\left(q\right)$.
To start we define,
\begin{equation}
\rho_\lambda = e^{-\beta \left({\cal H} - \lambda \Delta\right)}/\textrm{Tr}\left[e^{-\beta \left({\cal H} + \lambda \Delta\right)}\right]
\end{equation}
where,
\begin{equation}
\Delta = -\frac{1}{\beta}\log\left[\rho\right]-{\cal H}.
\end{equation}
We see that $\rho_\lambda = \rho_0$ if $\lambda = 0$ and $\rho_\lambda = \rho$ if $\lambda = 1$. Accordingly,
\begin{equation}
A\left[\rho\right] - A\left[\rho_0\right] = \int_0^1 \frac{\partial}{\partial \lambda} A\left[\rho_\lambda\right]\textrm{d}\lambda
\end{equation}
by the fundamental theorem of calculus. To evaluate the derivative we use,
\begin{eqnarray}
A\left[\rho_\lambda\right] & = & \\
& & \textrm{Tr}\left[\rho_\lambda\left({\cal H} + \lambda \Delta + \frac{1}{\beta}\log \left[\rho_\lambda\right]\right)\right]-\lambda \textrm{Tr}\left[\Delta\rho_\lambda\right].\nonumber
\end{eqnarray}
The first trace is stationary for variations of $\rho_\lambda$ about the corresponding density matrix. Thus, we only need to differentiate the second trace,
\begin{equation}
\frac{\partial}{\partial \lambda} A\left[\rho_\lambda\right] = -\lambda \textrm{Tr}\left[\Delta \frac{\partial}{\partial \lambda}\rho_\lambda\right].
\end{equation}
We evaluate  $\frac{\partial}{\partial \lambda}\rho_\lambda$ using the operator identity,
\begin{eqnarray}
\frac{\partial}{\partial \lambda} e^{-\beta \left({\cal H} + \lambda \Delta\right)} & = & e^{-\beta \left({\cal H} + \lambda \Delta\right)}\int_0^\beta\textrm{d}\beta'\Delta_\lambda\!\left(\beta'\right),\\
\Delta_\lambda\!\left(\beta'\right) & = & e^{\beta' \left({\cal H} + \lambda \Delta\right)} \Delta e^{-\beta' \left({\cal H} + \lambda \Delta\right)}\\
\frac{\partial}{\partial \lambda} \rho_\lambda & = & -\int_0^\beta\textrm{d}\beta' \rho_\lambda \left[\Delta_\lambda\!\left(\beta'\right)-\langle \Delta \rangle_\lambda\right],
\end{eqnarray}
where
\begin{equation}
\langle X \rangle_\lambda = \textrm{Tr}\left[\rho_\lambda X\right].
\end{equation}
Therefore,
\begin{equation}
\label{A10}
\frac{\partial}{\partial \lambda} A\left[\rho_\lambda\right] = \lambda \int_0^\beta \textrm{d}\beta' \left(\langle \Delta \Delta_\lambda\!\left(\beta'\right)\rangle_\lambda-\langle\Delta\rangle_\lambda^2\right).
\end{equation}
By cyclically permuting operators within the trace, one can verify that
\begin{eqnarray}
\langle \Delta_\lambda\left(\beta'\right)\rangle_\lambda & = & \langle \Delta \rangle_\lambda \,\,\,  \forall \beta',\\
\left\langle\Delta\Delta_\lambda\left(\beta'\right)\right\rangle_\lambda & = & \left\langle \Delta_\lambda\left(\frac{1}{2}\beta'\right)^\dagger \Delta_\lambda \left(\frac{1}{2}\beta'\right)\right\rangle.
\end{eqnarray}
With these identities, we can rewrite Eq.~\ref{A10},
\begin{align}
& \frac{\partial}{\partial \lambda} A\left[\rho_\lambda\right] = \\
& \lambda \int_0^\beta \!\textrm{d}\beta' \left\langle \!\!\left(\Delta_\lambda\left(\frac{1}{2}\beta'\right)-\left\langle\Delta\right\rangle_\lambda\right)^\dagger  \!\!\left(\Delta_\lambda\left(\frac{1}{2}\beta'\right)-\langle\Delta\rangle_\lambda\right)\!\right\rangle_\lambda\!\!\!.\nonumber
\end{align}
This integral is non-negative and can be zero only if $\Delta$ is a multiple of the unit operator, i.e., if $\rho_0 = \rho$. This proves that the minimum of the free energy must occur when $\rho_\lambda = \rho_0$.

\subsection{Proof of classical Bogoliubov inequality}

If $\eta_0\!\left(q\right)$ is the equilibrium density for a classical canonical ensemble and $\tilde{\eta_0}\!\left(q\right)$ is a different density, Gibbs' classical Bogoliubov inequality states that,
\begin{equation}
A\left[\tilde{\eta_0}\!\left(q\right)\right] > A\left[\eta_0\!\left(q\right)\right], \, \, \, \tilde{\eta_0}\!\left(q\right) \neq \eta_0\!\left(q\right)
\end{equation}
where $A$ is the classical Helmholtz free energy,
\begin{eqnarray}
A\left[\eta_0\!\left(q\right)\right] & = & E\left[\eta_0\!\left(q\right)\right]-\frac{1}{\beta}S\left[\eta_0\!\left(q\right)\right]\\
& = & \int \textrm{d}q \,\eta_0\!\left(q\right) W\!\left(q\right) + \frac{1}{\beta}\int \textrm{d}q \,\eta_0\!\left(q\right) \log\left[\eta_0\!\left(q\right)\right]\nonumber.
\end{eqnarray}
To see that this is the case we start by writing,
\begin{equation}
\frac{1}{\beta} \int \textrm{d}q \,\tilde{\eta_0}\!\left(q\right) \log\left[\tilde{\eta_0}\!\left(q\right)\right] \geq \frac{1}{\beta} \int \textrm{d}q \,\tilde{\eta_0}\!\left(q\right) \log\left[\eta_0\!\left(q\right)\right].
\end{equation}
The difference between the right and left sides of this equation is,
\begin{align}
& \frac{1}{\beta} \int \textrm{d}q\left(\tilde{\eta_0}\!\left(q\right) \log\left[\tilde{\eta_0}\!\left(q\right)\right] - \tilde{\eta_0}\!\left(q\right) \log\left[\eta_0\!\left(q\right)\right]\right) \\
& = \frac{1}{\beta} \int\textrm{d}q\,\tilde{\eta_0}\left(q\right) \log\left[\frac{\tilde{\eta_0}\!\left(q\right)}{\eta_0\!\left(q\right)}\right].\nonumber
\end{align}
Because $\log\left[x\right] \geq 1 - \frac{1}{x}$ and we know that the densities are normalized,
\begin{equation}
\frac{1}{\beta}\int\textrm{d}q\,\tilde{\eta_0}\left(q\right) \log\left[\frac{\tilde{\eta_0}\!\left(q\right)}{\eta_0\!\left(q\right)}\right] \geq \frac{1}{\beta} \int\textrm{d}q \left[\tilde{\eta_0}\!\left(q\right) - \eta_0\!\left(q\right)\right] = 0.
\end{equation}
We can simplify this further to,
\begin{equation}
\left\langle \frac{1}{\beta} \log\left[\eta_0\!\left(q\right)\right]\right\rangle \geq \left\langle \frac{1}{\beta} \log\left[\tilde{\eta_0}\!\left(q\right)\right]\right\rangle.
\end{equation}
We know that,
\begin{eqnarray}
\eta_0\!\left(q\right) & = & \frac{e^{-\beta E\left(q\right)}}{Z}\\
\tilde{\eta_0}\!\left(q\right) & = & \frac{e^{-\beta \widetilde{E}\left(q\right)}}{\widetilde{Z}}
\end{eqnarray}
where $\widetilde{E}\!\left(q\right)$ and $\widetilde{Z}$ correspond to the energy and partition function associated with $\tilde{\eta_0}\!\left(q\right)$. Thus,
\begin{eqnarray}
\left\langle \frac{1}{\beta} \log\left[\frac{e^{-\beta \widetilde{E}\left(q\right)}}{\widetilde{Z}}\right]\right\rangle\ & \geq & \left\langle \frac{1}{\beta} \log\left[ \frac{e^{-\beta E\left(q\right)}}{Z}\right] \right\rangle\\
\left\langle -\widetilde{E}\!\left(q\right) - \frac{1}{\beta}\log\left[\widetilde{Z}\right]\right \rangle & \geq & \left\langle - E\!\left(q\right) - \frac{1}{\beta}\log\left[Z\right] \right \rangle.
\end{eqnarray}
We may safely assume that $ \langle \widetilde{E}\!\left(q\right) \rangle = \left \langle E\!\left(q\right) \right\rangle$ so using the definition of the Helmholtz free energy, $A \equiv -\frac{1}{\beta}\log\left[Z\right]$,
\begin{equation}
A\left[\tilde{\eta_0}\!\left(q\right)\right] > A\left[\eta_0\!\left(q\right)\right], \, \, \, \tilde{\eta_0}\!\left(q\right) \neq \eta_0\!\left(q\right).
\end{equation}

\subsection{Applying linear functor to Silvera-Goldman}

The matrix which was ultimately used to transform the Silvera-Goldman potential was obtained by fitting 1,000 random potentials with $B=50$ basis functions in the appropriate $Q$ regime. The Silvera-Goldman potential has the form,
\begin{eqnarray}
V\!\left(r\right) & = & \exp{\left[\alpha - \delta r - \gamma r^2\right]} \\
& & -\left(\frac{C_6}{r^6}+\frac{C_8}{r^8} + \frac{C_{10}}{r^{10}}\right)f_c\!\left(r\right) + \frac{C_9}{r^9}f_c\!\left(r\right)\nonumber
\end{eqnarray}
where
\begin{equation}
f_c\!\left(r\right)=\begin{cases}
e^{-\left(r_c/r - 1\right)^2}, & \text{if $r\leq r_c$}\\
1, & \text{otherwise}.
\end{cases}
\end{equation}
Parameters for the Silvera-Goldman potential are provided in Table 1  \cite{Silvera1978}.
\begin{table}[H]
\caption{Parameters of the Silvera-Goldman potential \cite{Silvera1978}.}
\begin{tabularx}{\columnwidth}{>{\centering}m{0.45\columnwidth} >{\centering}m{0.45\columnwidth}}
\hline\hline
\noalign{\vskip 1mm}    
Parameter & Value (atomic units) \tabularnewline
\noalign{\vskip 1mm}    
\hline
\noalign{\vskip 2mm}    
$\alpha$ & 1.713 \tabularnewline
$\delta$ & 1.5671 \tabularnewline
$\gamma$ & 0.00993 \tabularnewline
$C_6$ & 12.14 \tabularnewline
$C_8$ & 215.2 \tabularnewline
$C_9$ & 143.1 \tabularnewline
$C_{10}$ & 4813.9 \tabularnewline
$r_c$ & 8.321 \tabularnewline
\noalign{\vskip 1mm}    
\hline\hline
\end{tabularx}
\end{table}

Exponential functions cannot be easily represented in a polynomial basis and the Silvera-Goldman potential diverges exponential as $r$ approaches zero. Accordingly, we fit the potential only in the physically relevant region of $r > 4$ Bohr. We matched the slope of the potential at $r = 4$ Bohr and extend the potential as a straight line in the region $0 < r < 4$ Bohr. We choose to fit the potential out to $r = 24$ Bohr but imposed a standard cutoff after the fact at $r = 20$ Bohr as the potential is clearly flat by this point.  We simulated \emph{para}-hydrogen at 14K and 25K. At 25K, the thermal de Broglie wavelength is 4.6 Bohr; thus, a cutoff distance of 20 Bohr gives $Q = 0.23$. At 14K, the thermal de Broglie wavelength is 6.2 Bohr and $Q = 0.31$. Based on statistics collected from 10,000 random potentials generated with these $Q$ values, in both of these regimes, the classical free energy is more accurate than the ${\cal F}$ predicted free energy less than 1\% of the time.\newpage

\section{Acknowledgements}
The authors thank David Manolopoulos, Eugene Shakhnovich, and Eric Heller for helpful conceptual discussions regarding direction of the project. We also thank Jarrod McClean, Joseph Goodknight and Nicolas Sawaya for discussions regarding revision of the manuscript. Research sponsored by the United States Department of Defense. The views and conclusions contained in this document are those of the authors and should not be interpreted as representing the official policies, either expressly or implied, of the U.S. Government.

\bibliographystyle{ieeetr}
\bibliography{./library,Functor}

\section{Author contributions}
All authors conceived and designed the research project. With guidance from J.P. and A.A.G., R.B. wrote the proofs, first draft of the manuscript, and code for obtaining and characterizing the numerical functor. All authors interpreted the results and co-wrote the article. The authors declare no competing financial interests. Correspondence should be addressed to R.B. (babbush@fas.harvard.edu) and A.A.G. (aspuru@chemistry.harvard.edu).

\end{document}